\documentstyle[12pt]{article}
\def\simlt{\stackrel{<}{{}_\sim}}

\def\be{\begin{equation}}
\def\ee{\end{equation}}
\def\bear{\be\begin{array}}
\def\eear{\end{array}\ee}
\def\bea{\begin{eqnarray}}
\def\eea{\end{eqnarray}}

\def\baselinestretch{1}
\begin{document}
\catcode`@=11
\newtoks\@stequation
\def\subequations{\refstepcounter{equation}%
\edef\@savedequation{\the\c@equation}%
  \@stequation=\expandafter{\theequation}
  \edef\@savedtheequation{\the\@stequation}
  \edef\oldtheequation{\theequation}%
  \setcounter{equation}{0}%
  \def\theequation{\oldtheequation\alph{equation}}}
\def\endsubequations{\setcounter{equation}{\@savedequation}%
  \@stequation=\expandafter{\@savedtheequation}%
  \edef\theequation{\the\@stequation}\global\@ignoretrue

\noindent}
\catcode`@=12
\begin{titlepage}
\title{{\bf Problems for Supersymmetry Breaking by the Dilaton
            in Strings from Charge and Color Breaking}
\thanks{Research supported in part by: the CICYT, under
contracts AEN95-0195 (JAC) and AEN93-0673 (ALL, CM); the European Union,
under contracts CHRX-CT92-0004 (JAC), CHRX-CT93-0132 (CM) and
SC1-CT92-0792 (CM); the
Ministerio de Educaci\'on y Ciencia, under FPI grant (ALL).}
}
\author{ {\bf J.A. Casas\thanks{On leave of absence from Instituto de
Estructura de la Materia CSIC, Serrano 123, 28006 Madrid, Spain.}
${}^{ {\footnotesize,\S}}$},
{\bf A. Lleyda${}^{ {\footnotesize\P}}$ }
and {\bf C. Mu\~noz${}^{\footnotesize\P}$}\\
\hspace{3cm}\\
${}^{\footnotesize\S}$ {\small Santa Cruz Institute for Particle Physics}\\
{\small University of California, Santa Cruz, CA 95064, USA}\\
{\small casas@cc.csic.es}\\
\vspace{-0.3cm}\\
${}^{\footnotesize\P}$ {\small Departamento de F\'{\i}sica
Te\'orica C--XI} \\
{\small Universidad Aut\'onoma de Madrid, 28049 Madrid, Spain}\\
{\small amanda@delta.ft.uam.es $$ cmunoz@ccuam3.sdi.uam.es}}
\date{}
\maketitle
\def\baselinestretch{1.15}
\begin{abstract}
\noindent
The general constraints on the parameter space of soft-breaking terms, in order
to avoid dangerous charge and color breaking minima, are applied to the
four-dimensional string scenario where the dilaton is the source of
supersymmetry breaking (dilaton-dominated limit).
The results indicate that the whole parameter space is excluded on these
grounds after imposing the present experimental data on the top mass.
The inclusion of a non-vanishing cosmological constant does not improve
essentially the prospects.
Possible way-outs to this situation are briefly discussed.
\end{abstract}

\thispagestyle{empty}

\leftline{}
\leftline{}
\leftline{FTUAM 96/03}
\leftline{SCIPP-96-03}
\leftline{IEM-FT-124/96}
\leftline{hep-ph/9601357}
\leftline{January 1996}
\leftline{}

\vskip-23.cm
\rightline{}
\rightline{ FTUAM 96/03}
\rightline{SCIPP-96-03}
\rightline{ IEM-FT-124/96}
\rightline{hep-ph/9601357}
\vskip3in

\end{titlepage}
\newpage
\setcounter{page}{1}

\section{Introduction}

The presence of scalar fields with color and electric charge in
supersymmetric (SUSY) theories induces the possible existence of
dangerous charge and color breaking minima, which would make
the standard vacuum unstable [1-8]. 
This is not necessarily a shortcoming since many SUSY models can
be discarded on these grounds, thus improving the predictive power
of the theory. A complete
analysis of all the potentially dangerous directions in the field
space of the minimal supersymmetric standard model (MSSM) was carried
out in ref.\cite{CCB}. It was shown there that the corresponding constraints
on the soft parameter space ($m$, $M$, $A$, $B$) are very strong.
As a matter of fact, there are extensive regions of this space that
{\it become forbidden} producing
important bounds, not only on the value of the trilinear scalar term
($A$), but also on the values of the bilinear scalar term ($B$) and the
scalar and gaugino masses ($m, M$ respectively).

On the other hand, in four-dimensional strings, working at the perturbative
level, it is possible to obtain
information about the structure of soft SUSY-breaking terms
[9-14]. The basic idea is to identify some chiral fields whose auxiliary 
components could break SUSY by acquiring a vacuum expectation value 
(VEV). This is the case of the dilaton and the moduli fields.  
The important point in this assumption of locating the seed of 
SUSY breaking in the dilaton/moduli sector, is that it leads to some 
interesting relationships among different soft terms which could 
perhaps be experimentally tested.  
This general analysis was applied in particular to the gaugino
condensation scenario in ref.\cite{gaugino}, whereas in refs.[10-14] 
no special assumption was made about the possible origin of 
SUSY breaking.


The dilaton-dominated limit \cite{Kaplunovsky, Brignole}, where only the 
dilaton field, $S$, contributes to SUSY breaking\footnote{ For possible 
explicit SUSY-breaking mechanisms where this limit might be obtained see
ref.\cite{Halyo}.} is specially interesting.
The dilaton field, whose VEV determines the tree-level gauge coupling, is
present in any four-dimensional string and couples at tree-level in
a universal manner to all particles.
Therefore, this limit is {\it model independent} and, as a consequence,
the soft terms are independent of the four-dimensional string
considered.
Besides, their expressions are quite simple since they are universal and
essentially depend on only two parameters, $m_{3/2}$ and $B$,
where $m_{3/2}$ is the gravitino mass.
Actually, universality is a desirable property, not only to reduce the
number of independent parameters in the MSSM, but also for
phenomenological reasons, particularly to avoid flavour changing
neutral currents. Because of the simplicity of this scenario, the
corresponding low-energy predictions
are quite precise \cite{Barbieri, Brignole, Vissani}.
For example, the first and second generation squarks are almost
degenerate with the gluino and are much heavier than sleptons\footnote{
The phenomenology of SUSY breaking by the dilaton in the context of a
flipped SU(5) model was also studied in ref.\cite{Lopez}.}.

{}From all the above reasons it is clearly of the utmost importance
to study the consistency of the dilaton-dominated scenario
with the possible existence of dangerous charge and color breaking minima.
This is the aim of this paper.
In fact, we will show that charge and color breaking
constraints are so important that
the whole parameter space is {\it forbidden} and, as a consequence, the
dilaton-dominated limit is excluded on these grounds.

\section{Basic Ingredients}
Let us briefly review the basic ingredients required for this analysis.
First we will concentrate on the form of soft breaking terms. The
general form of the soft SUSY-breaking Lagrangian in the context of the
MSSM is given by
\bea
\label{lsoft}
{\cal L}_{soft}=\frac{1}{2} \sum_{a=1}^{3}
M_a {\overline{\lambda}}_a \lambda_a  - \sum_{i} m_i^2 |\phi_i|^2
-(A_{ijk} W_{ijk} + B \mu H_1 H_2 + {\rm h.c.}),
\eea
where $W_{ijk}$ are the usual terms of the Yukawa superpotential of the
MSSM with \linebreak 
$i$=$Q_L, u_L^c, d_L^c, L_L, e_L^c, H_1, H_2,$ and $\phi_i,\lambda_a$
are the canonically normalized scalar and gaugino fields respectively.
In the dilaton-dominated scenario \cite{ Kaplunovsky, Brignole},
neglecting string loop corrections, one obtains the following results for
the above scalar masses, gaugino masses and soft trilinear terms
\bea
\label{softtermsV}
m_i^2   &=& m_{3/2}^2 +V_0                               \;\;,\nonumber\\
M_a     &=&   \sqrt{3 m_{3/2}^2 +V_0} \;\; e^{-i \alpha} \;\;,\nonumber\\
A_{ijk} &=& - \sqrt{3 m_{3/2}^2 +V_0} \;\; e^{-i \alpha} \;\;,
\eea
where, for the sake of completeness, we have included
the VEV of the scalar potential (i.e. the
cosmological constant) $V_0$, and a possible phase $\alpha$ of the
dilaton F-term \cite{Brignole}.
Notice that we are using the standard supergravity mass units where
$M_{Planck}/{\sqrt{8 \pi}}$=$1$.

The value of the bilinear term $B$ is more model dependent and deserves
some additional comments. Indeed, $B$ depends not only on the 
dilaton-dominance assumption but also on the particular mechanism which could
generate the associated (electroweak size) $\mu$ term \cite{Munoz1}. 
For example, the interesting possibilities of generating it through the
K\"ahler potential \cite{Giudice, Casas, Kaplunovsky, Lopes, Antoniadis} 
or the superpotential \cite{Casas, Antoniadis} give rise, 
in the dilaton-dominated limit, to the following value 
of $B$ \cite{Brignole, Kaplunovsky, Munoz2, Munoz1, Brignole2}:
%
%
\bea
\label{BV}
B = 2 m_{3/2} + \frac{V_0}{m_{3/2}} \;\;\;.
\eea
The previous expressions for the soft terms can be simplified taking
into account several experimental restrictions.  From the limits on
the electric dipole moment of the neutron it seems reasonable to
impose in what follows $\alpha$=$0$ mod $\pi$.  On the other hand,
experimental constraints in present cosmology allow us to assume
vanishing cosmological constant $V_0$=$0$ (we will see later on that our
conclusions will not be modified if we give up this assumption). Then
%
\bea
\label{softterms}
m_i^2 &=& m_{3/2}^2 \;\; ,\nonumber\\ M_a &=& \pm \sqrt{3} \; m_{3/2}
\;\; ,\nonumber\\ A_{ijk} &=& - M_a \;\;,
\eea
and the $B$ term associated with the mechanisms explained above in
order to solve the $\mu$ problem is\footnote{Phenomenological aspects of the 
$B$=$2m_{3/2}$ scenario have been studied in refs.\cite{ Barbieri, Brignole,
MAD}.}
\bea
\label{B}
B = 2 m_{3/2} \;\;\;.
\eea
Expressions (\ref{softtermsV}--\ref{B}) are to be understood at the string
scale $M_{string}\simeq 0.5\times g_{string}\times{10}^{18}$ GeV
\cite{Kaplu},
where $g_{string}=(Re\; S)^{-1/2}\simeq 0.7$. In the following we will
assume that
the discrepancy between the unification
scale of the gauge couplings, $M_X$, and the string unification scale,
$M_{string}$, can be explained by the effect of string threshold
corrections \cite{Ross}.

In the present paper we have taken the expressions of the soft
terms given by eqs.(\ref{softterms}, \ref{B}) as our starting point to work,
considering the value of $B$ given by eq.(\ref{B}) as a guiding example.
Then, by varying the value of $B$ (and also $V_0$) we will eventually obtain
the most general results.
Concerning the value of the $\mu$ parameter, we
will fix it as usual from the requirement of correct
electroweak breaking\footnote{The value of $\mu$ can also be fixed once
we choose a particular mechanism for solving the $\mu$ problem, e.g.
if $\mu$ is generated through the K\"ahler
potential in dilaton-dominated orbifold models, then
$\mu$=$m_{3/2}$ \cite{Brignole2, Munoz3}.
However, we prefer to eliminate $\mu$ in terms of
the other parameters by imposing appropriate electroweak breaking,
as mentioned above, the reason being that this provides the most
general analysis.}.

\vspace{0.3cm}
\noindent
The second basic ingredient of our analysis concerns the constraints
associated with the existence of dangerous directions in the field space.
As was mentioned in the
introduction, a complete analysis of this issue,
including
in a proper way the radiative corrections to the
scalar potential, was carried out in ref.\cite{CCB}.
The most relevant results obtained there for our present task
are the following.

There are two types of constraints:
the ones arising from directions in the field-space along
which the (tree-level) potential can become unbounded from below (UFB),
and those arising from the existence of charge and color
breaking (CCB) minima in the potential deeper than the
standard minimum.

Concerning the UFB directions (and corresponding constraints),
there are three of them, labelled as UFB-1, UFB-2, UFB-3
in \cite{CCB}. It is worth mentioning here that in general the
unboundedness is only true
at tree-level since radiative corrections eventually raise the potential for
large enough values of the fields, but still these minima can be deeper than
the realistic one (i.e. the SUSY standard model vacuum) and thus dangerous.
The UFB-3 direction, which involves
the fields
$\{H_2,\nu_{L_i},e_{L_j},e_{R_j}\}$ with $i \neq j$
and thus leads also to electric charge
breaking, yields the {\it strongest} bound among {\it all}
the UFB and CCB constraints. The explicit form of this bound
is as follows.
By simple analytical minimization it is possible to write the
value of all the relevant fields along the UFB-3 direction in
terms of the $H_2$ one. Then, for any value of $|H_2|<M_{string}$ satisfying
\be
\label{SU6}
|H_2| > \sqrt{ \frac{\mu^2}{4\lambda_{e_j}^2}
+ \frac{4m_{L_i}^2}{g'^2+g_2^2}}-\frac{|\mu|}{2\lambda_{e_j}} \ ,
\ee
the value of the potential along the UFB-3 direction is simply given
by
\be
\label{SU8}
V_{\rm UFB-3}=(m_2^2 -\mu^2+ m_{L_i}^2 )|H_2|^2
+ \frac{|\mu|}{\lambda_{e_j}} ( m_{L_j}^2+m_{e_j}^2+m_{L_i}^2 ) |H_2|
-\frac{2m_{L_i}^4}{g'^2+g_2^2} \ .
\ee
Otherwise
\be
\label{SU9}
V_{\rm UFB-3}= (m_2^2 -\mu^2 ) |H_2|^2
+ \frac{|\mu|} {\lambda_{e_j}} ( m_{L_j}^2+m_{e_j}^2 ) |H_2| + \frac{1}{8}
(g'^2+g_2^2)\left[ |H_2|^2+\frac{|\mu|}{\lambda_{e_j}}|H_2|\right]^2 \ .
\ee
In eqs.(\ref{SU8},\ref{SU9}) $\lambda_{e_j}$ is the leptonic Yukawa
coupling of the $j-$generation and $m_2^2$ is the sum of the $H_2$ squared
soft mass, $m_{H_2}^2$, plus $\mu^2$. Then, the
UFB-3 condition reads
\be
\label{SU7}
V_{\rm UFB-3}(Q=\hat Q) > V_{\rm real \; min} \ ,
\ee
where $V_{\rm real \; min}=-\frac{1}{8}\left(g'^2 + g_2^2\right)
\left(v_2^2-v_1^2\right)^2$, with $v_{1,2}$ the VEVs of the Higgses $H_{1,2}$,
is the realistic minimum evaluated at $M_S$ (see below)
and the $\hat Q$ scale is given by \linebreak
$\hat Q\sim {\rm Max}(g_2 |e|, \lambda_{top} |H_2|,
g_2 |H_2|, g_2 |L_i|, M_S)$
with
$|e|$=$\sqrt{\frac{|\mu|}{\lambda_{e_j}}|H_2|}$ and
$|L_i|^2$=$-\frac{4m_{L_i}^2}{g'^2+g_2^2}$ \linebreak +($|H_2|^2$+$|e|^2$).
Finally, $M_S$ is the typical scale of SUSY masses (normally a good
choice for $M_S$ is an average of the stop masses, for more details
see refs.\cite{Gamberini, bea, CCB}).
Notice from (\ref{SU8}-\ref{SU9}) that the negative contribution to $V_{UFB-3}$
is essentially given by the $m_2^2-\mu^2$ term, which can be very sizeable in 
many instances. On the other hand, the positive contribution is dominated by 
the term $\propto 1/\lambda_{e_j}$, thus the larger
$\lambda_{e_j}$ the more restrictive
the constraint becomes. Consequently, the optimum choice of
the $e$--type slepton is the third generation one, i.e.
${e_j}=$ stau.

Concerning the CCB constraints, let us mention that the ``traditional'' CCB
bounds \cite{Frere}, when correctly evaluated (i.e. including the
radiative corrections in a proper way), turn out to be extremely weak.
However, the ``improved" set of analytic constraints obtained in
ref.\cite{CCB}, which represent the
necessary and sufficient conditions to avoid dangerous CCB minima,
is much stronger. It is not possible to give here an account of the
explicit form of the CCB constraints used in the present paper. This
can be found in section 5 of ref.\cite{CCB}, to which we refer the
interested reader.

\section{Results}

In Fig.1 we have presented in detail the interesting case (and guiding
example) $B$=$2m_{3/2}$ for the two possible values of gaugino masses,
$M \equiv M_a$=$\; \pm \sqrt{3} m_{3/2}$, see eqs.(\ref{softterms},
\ref{B}).
In the plots of the figure we have shown how the parameter space defined
by $m_{3/2}$ (the only soft parameter left independent) and $M_{top}$
(which we let vary for completeness) is totally excluded by the different
constraints in the game. Some comments are in order here.

First, we have taken $m_{3/2}\leq 500$ GeV since
larger values of $m_{3/2}$ would induce too large SUSY-mass spectra; e.g.
$m_{3/2}$=500 GeV implies gluino and first and second generation squark
masses of order 2.5 TeV, conflicting
the absence-of-fine-tuning requirements \cite{bar, bea}. On the other hand,
rather than assuming a particular value of the top mass, we have preferred
to vary the physical (pole) top mass, $M_{top}^{phys}$, between 150 and 200
GeV.
Actually, it is not always possible to choose the boundary condition of the
top Yukawa coupling $\lambda_{top}$ so that the physical (pole) mass is
reproduced because the renormalization group (RG) infrared fixed point of
$\lambda_{top}$ puts an
upper bound on the running top mass $M_{top}$, namely
$M_{top}\simlt 197 sin\beta$
GeV \cite{Inoue}, where $tan \beta$=$ v_2/v_1$.
The corresponding restriction in the parameter space (black regions in Fig.1)
is certainly substantial
in the case
$M$=$\sqrt{3} m_{3/2}$, yielding $M_{top}^{phys}<167$ GeV, which is by itself
a remarkable result\footnote{However, in the case $M$=$-\sqrt{3} m_{3/2}$ the
restriction is small allowing in principle large top masses. This possibility
was not taken into account in ref.\cite{Barbieri} where an upper bound of
180 GeV on the top mass was obtained for $M$=$\sqrt{3}m_{3/2}$,
$A$=$-\sqrt{3}m_{3/2}$. The discrepancy between that upper bound and the one
that we obtain here for the same case, 167 GeV, is due to the following:
the RGEs \cite{Ibanez} used in ref.\cite{Barbieri}
correspond to the soft Lagrangian of eq.(\ref{lsoft}) with a minus sign
in front of the gaugino masses and therefore the associated boundary
conditions that
should have been used in order to get the correct result
are $M$=$-\sqrt{3}m_{3/2}$, $A$=$-\sqrt{3}m_{3/2}$.}.

The region excluded by the CCB bounds is denoted by circles in the figure.
The sign of the trilinear soft term $A_{ijk}$ is important in these type of
constraints as can be seen in the figure (recall that
$A\equiv A_{ijk}$=$-M$).
Whereas the case $M$=$-\sqrt{3}m_{3/2}$ is not constrained at all, in the
case $M$=$\sqrt{3}m_{3/2}$ the whole parameter space left allowed by the
previous ``top-fixed-point
constraint'' is excluded by the CCB bounds.

Anyway, it is apparent from Fig.1 that the
the restrictions coming from the UFB constraints (small filled squares)
are very strong in both cases.
Most of the parameter space is in fact excluded by the UFB-3
constraint, which has been explained in the previous section.

Finally, we have also plotted in Fig.1 the region excluded
by the experimental bounds on SUSY particle masses (filled diamonds).
Conservatively enough, we have imposed
\bea
\label{experimentalb}
& &M_{\tilde g} \geq 120\ {\rm GeV}        \;,\;\;
   M_{\tilde \chi^{\pm}}\geq 45\ {\rm GeV} \;,\;\;
   M_{\tilde \chi^o} \geq 18\ {\rm GeV}    \;,\nonumber \\
& &M_{\tilde q}\geq 100\ {\rm GeV}  \;,\;\;\;
   M_{\tilde t} \; \geq 45\ {\rm GeV}\;,\;\;\; \;
   M_{\tilde l} \; \geq 45\ {\rm GeV}\; ,
\eea
in an obvious notation.
The corresponding forbidden area comes mainly from too small masses for
neutralinos and charginos in the case $M$=$\sqrt{3} m_{3/2}$, and
for the sleptons when $M$=$-\sqrt{3} m_{3/2}$.
The ants indicate regions which are excluded by negative squared mass
eigenvalues, in this case the sneutrinos.

Notice from Fig.1 that there are areas that are simultaneously constrained
by different types of bounds.

At the end of the day, the allowed region (white) left is very small.
Only in the case $M$=$-\sqrt{3}m_{3/2}$ and for $m_{3/2}$ larger than 320
GeV, which on the other hand corresponds to gluino and first and second
generation squark masses heavier than 1.5 TeV, the dangerous minima are
not present.
However, this occurs for $M_{top}^{phys} <$ 157 GeV. Thus, using the present
experimental data on the top mass,
$M_{top}^{exp}$=$180 \pm 12$ GeV \cite{PDG}, we conclude that
the dilaton dominance limit for SUSY breaking in strings with $B$=$2m_{3/2}$
is excluded on charge and color breaking (CCB and UFB) grounds.
It is worth mentioning here that, even if we relax
the previous fine-tuning requirement (i.e. $m_{3/2}\leq 500$ GeV) by 
admitting higher values of $m_{3/2}$, we have checked that the growing
of the allowed region is anyway remarkably slow, so one has to go to
extremely high values of $m_{3/2}$ to get acceptable top masses.

\vspace{0.3cm}
\noindent
Let us generalize now the previous analysis by varying the value of $B$.
In Fig.2  we have shown the representative examples $B$=$0, 3m_{3/2}$ with
$M$=$-\sqrt{3}m_{3/2}$. Whereas for large values of $B$, $B\geq 3m_{3/2}$,
the whole parameter space is
excluded, for $B$=$0$ there is still a very small allowed region.
However, this region is in fact excluded by the value of
$M_{top}^{exp}$ in a similar fashion as it happened in the $B=2m_{3/2}$
case (see above).
Intermediate values of $B$
do not improve the situation.
For negative values of $B$ the corresponding figures are the same, since
they are invariant under the transformation $B$, $A$, $M\rightarrow$
-$B$, -$A$, -$M$.
The same conclusion is obtained for the case $M$=$\sqrt{3}m_{3/2}$.
Examples of this scenario are given in Fig.1 for $B$=$2m_{3/2}$ and
in Fig.2 for $B$=$0$ (notice that from the previous invariance the
$B$=$0$ figure is valid for $M$=$-\sqrt{3}m_{3/2}$ as well as for
$M$=$\sqrt{3}m_{3/2}$).


{}From the various figures it is clear that the CCB and UFB constraints
do not
allow the possibility of SUSY breaking in the dilaton-dominance
limit in strings.

Let us now consider for the sake of completeness the possibility of a
non-vanishing cosmological constant $V_0$.
This may correspond to different attitudes concerning this problem.
For example, one might think that the experimentally constrained cosmological
term in present cosmology is not directly connected to the particle physics
vacuum energy $V_0$.
Another possibility is to admit a non-vanishing tree-level cosmological
constant, requiring a vanishing fully renormalized one \cite{Choi}.
Anyway, whatever the assumption is, it is worth relaxing the condition
$V_0$=$0$ since, from eq.(\ref{softtermsV}), this might be the only
possibility to avoid the previous dramatical conclusions for the
dilaton-dominated limit. Notice, however, that $V_0$ is constrained to be
$V_0\ge -m_{3/2}^2$ in order to
avoid negative squared scalar masses (see eq.(\ref{softtermsV})) and, on the
other hand, it should not be
too large in order to keep the SUSY spectrum in the range of 1 TeV.
In addition, to simplify the analysis we have initially taken the
(theoretically well-motivated) value of $B$ given in eq.(\ref{BV}).

In Fig.3 we have presented three representative examples of a non-vanishing
cosmological constant ($V_0$=$-0.5 m_{3/2}^{2},\; 3 m_{3/2}^{2},
\;10 m_{3/2}^{2}$) with $M$=$-\sqrt{3}m_{3/2}$.
We see that for $V_0$=$-0.5m_{3/2}^2$ the whole parameter space is
excluded.
Larger values of $V_0$ do improve the situation, the best case being
$V_0$=$3m_{3/2}^2$.
However, once more, the present value of $M_{top}^{exp}$ essentially excludes
this possibility for any reasonable SUSY spectrum. More precisely,
notice that for $m_{3/2}\geq 300$ GeV we are already in the range of 2 TeV
for the gluino and first and second generation squark masses, which are far
too large from fine-tuning considerations.
Finally, for the case $M$=$\sqrt{3}m_{3/2}$ (not represented in the
figures) the constraints are even stronger,
e.g. for $V_0$=$3m_{3/2}^2$ the whole parameter space is excluded.

Even if we let $B$ vary as an independent parameter (at the same time
as $V_0$) the results do not improve much. More precisely, the best case
is found for $V_0$=$3m_{3/2}^2$, $B$=$3m_{3/2}$. But, even in this extreme
possibility, if one demands a reasonable SUSY spectrum (e.g. masses of
order 1 TeV, which in this case requires $m_{3/2}\sim $ 150 GeV) the
top mass becomes too small ($M_{top}^{phys}\leq$ 172 GeV), almost inconsistent
with the present experimental lower bound ($M_{top}^{exp}\geq$ 168 GeV).

To summarize the results, the dilaton-dominated limit is essentially
excluded on charge and color breaking (CCB and UFB) grounds.
Even in the few extreme cases consistent with
the charge and color breaking constraints, the spectrum is inviable since
either the top mass
is too small or the SUSY spectrum is far too heavy from any fine-tuning
consideration. The addition of a non-vanishing cosmological constant
in the game does not improve this situation.

\section{Concluding Remarks}
The dilaton-dominated limit in strings, defined as the scenario where
the only source of SUSY breaking is the dilaton, is highly interesting
since at the string tree-level approximation its formulation is model
independent (i.e. holds for any four-dimensional string) and yields
quite precise low-energy predictions (SUSY spectrum). However, we have seen
along this paper that, after imposing the present experimental 
data on the top mass, the whole parameter space of this scenario 
($m_{3/2}, B$) is
excluded on charge and color breaking grounds \cite{CCB}, i.e. by the
existence of charge and color breaking
minima deeper than the standard vacuum. Even allowing a non-vanishing
cosmological constant does not improve essentially the situation.
Due to the attractiveness of
the dilaton-dominated limit, let us briefly discuss some possible way-outs
to the previous dramatical conclusions.

One possibility is to accept that we live in a metastable vacuum, provided
that its lifetime is
longer than the present age of the universe \cite{Claudson, Riotto}, thus
rescuing some points in the parameter space. This possibility, however,
poses the cosmological problem of why our universe does not correspond
to the global minimum of the potential (without invoking an anthropic
principle). Even if a solution to that problem is found we would still
have to face the rather bizarre situation of a future cosmological
catastrophe, which does not seem very attractive. In addition, it is hard
to understand how  the cosmological constant is vanishing precisely
in such ``interim'' vacuum. Anyway, despite the previous shortcomings,
this is still a possible scenario which would be worth analyzing 
\cite{Nosotros}.

A different possibility is to assume that also the moduli fields
contribute to SUSY breaking.
Then the soft terms are modified and possibly some regions in the
parameter space would be allowed.
This more general situation deserves further analysis \cite{Nosotros}. 
Of course, this amounts to a departure of the pure dilaton-dominated
scenario. On the other hand, it is interesting to note that explicit
possible scenarios of SUSY breaking by gaugino condensation
in strings, when analyzed at the one-loop level, lead to the mandatory
inclusion of the moduli in the game 
(in fact the moduli are the main source of SUSY breaking in these cases)
\cite{gaugino, gaugino2}.

Finally, one may think that the perturbative and non-perturbative corrections
to the ``standard'' tree-level dilaton-dominated scenario are important
and can modify the previous conclusions. Due to analyticity and
non-renormalization theorems these contributions are likely to affect in
a substantial way only the K\"ahler potential \cite{Banksdine}
(in fact, the gauge kinetic function does also receive perturbative
corrections at one loop level, but not beyond). 
Actually,
the one-loop string corrections to the K\"ahler potential (and the gauge 
kinetic function) have been
calculated for orbifold models \cite{Kaplder} and they are rather small
for sensible values of the moduli. Thus, it is reasonable to expect
that further perturbative corrections will be even smaller. However,
this is not the case for the string non-perturbative corrections, whose
size could be much larger (see e.g. ref.\cite{Shenker}). These 
corrections
could be crucial to understand both the SUSY-breaking mechanism and the
vanishing of the cosmological constant \cite{Banksdine}; actually, it 
is possible to show \cite{Progress} that a tree-level dilaton-dominated 
scenario cannot have a global minimum of the dilaton potential at vanishing
cosmological constant. Unfortunately, the form of this type of corrections
is very poorly known, which introduces additional sources of uncertainty
in the analysis \cite{Banksdine, Progress}.

\section*{Acknowledgements}
We thank L.E. Ib\'a\~nez for useful comments.



\def\MPL #1 #2 #3 {{\em Mod.~Phys.~Lett.}~{\bf#1}\ (#2) #3 }
\def\NPB #1 #2 #3 {{\em Nucl.~Phys.}~{\bf B#1}\ (#2) #3 }
\def\PLB #1 #2 #3 {{\em Phys.~Lett.}~{\bf B#1}\ (#2) #3 }
\def\PR  #1 #2 #3 {{\em Phys.~Rep.}~{\bf#1}\ (#2) #3 }
\def\PRD #1 #2 #3 {{\em Phys.~Rev.}~{\bf D#1}\ (#2) #3 }
\def\PRL #1 #2 #3 {{\em Phys.~Rev.~Lett.}~{\bf#1}\ (#2) #3 }
\def\PTP #1 #2 #3 {{\em Prog.~Theor.~Phys.}~{\bf#1}\ (#2) #3 }
\def\RMP #1 #2 #3 {{\em Rev.~Mod.~Phys.}~{\bf#1}\ (#2) #3 }
\def\ZPC #1 #2 #3 {{\em Z.~Phys.}~{\bf C#1}\ (#2) #3 }

\newpage

\section*{Figure Captions}

\begin{description}
\item[Fig.1] Excluded regions in the parameter space of the MSSM
assuming SUSY breaking by the dilaton, with $B$=$2m_{3/2}$.
The black region is excluded because it is not possible to reproduce the
experimental mass of the top.
The small filled squares indicate regions excluded by Unbounded {}From Below
constraints.
The circles indicate regions excluded by Charge and Color Breaking
constraints.
The filled diamonds correspond to regions excluded by the experimental lower
bounds on SUSY-particle masses.
The ants indicate regions excluded by negative scalar squared mass eigenvalues.
\item[Fig.2] The same as Fig.1 but with $B$=$0,\;3m_{3/2}$ and
$M$=$-\sqrt{3} m_{3/2}$.
\item[Fig.3] The same as Fig.1 but with $M$=$-\sqrt{3} m_{3/2}\;$ and
$\;V_0$=$-0.5 m_{3/2}^{2},\;3 m_{3/2}^{2},\;10 m_{3/2}^{2}$.

\end{description}

%
\input{psfig.tex}
\newpage
\thispagestyle{empty}
\psfig{file=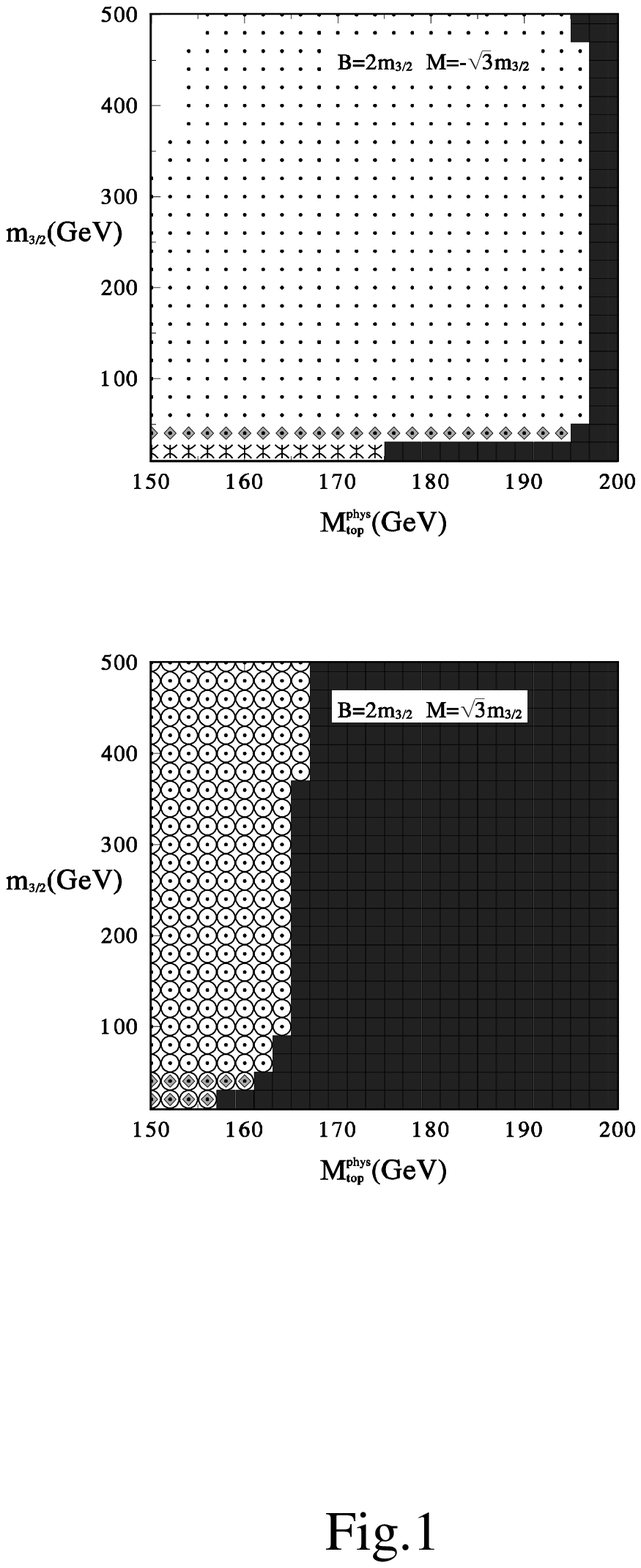,height=23.1cm,angle=180}
\newpage
\thispagestyle{empty}
\psfig{file=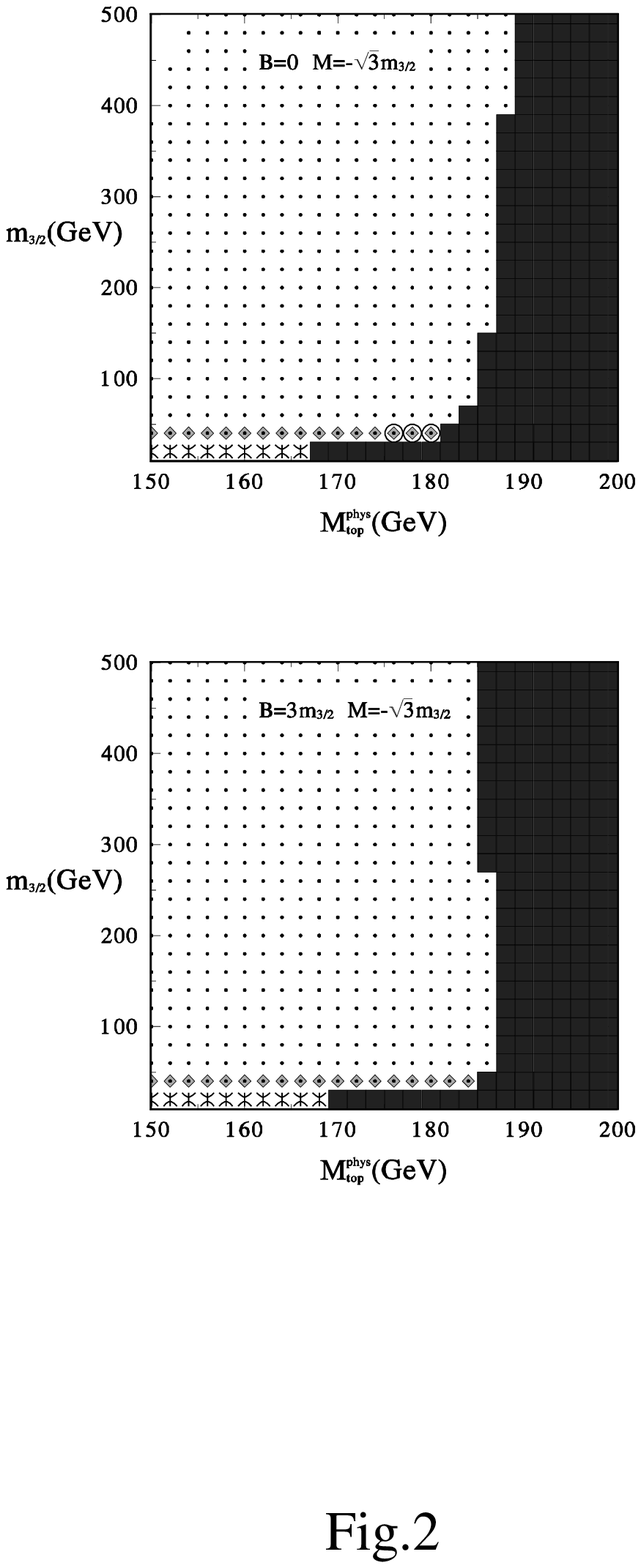,height=23.1cm,angle=180}
\newpage
\thispagestyle{empty}
\psfig{file=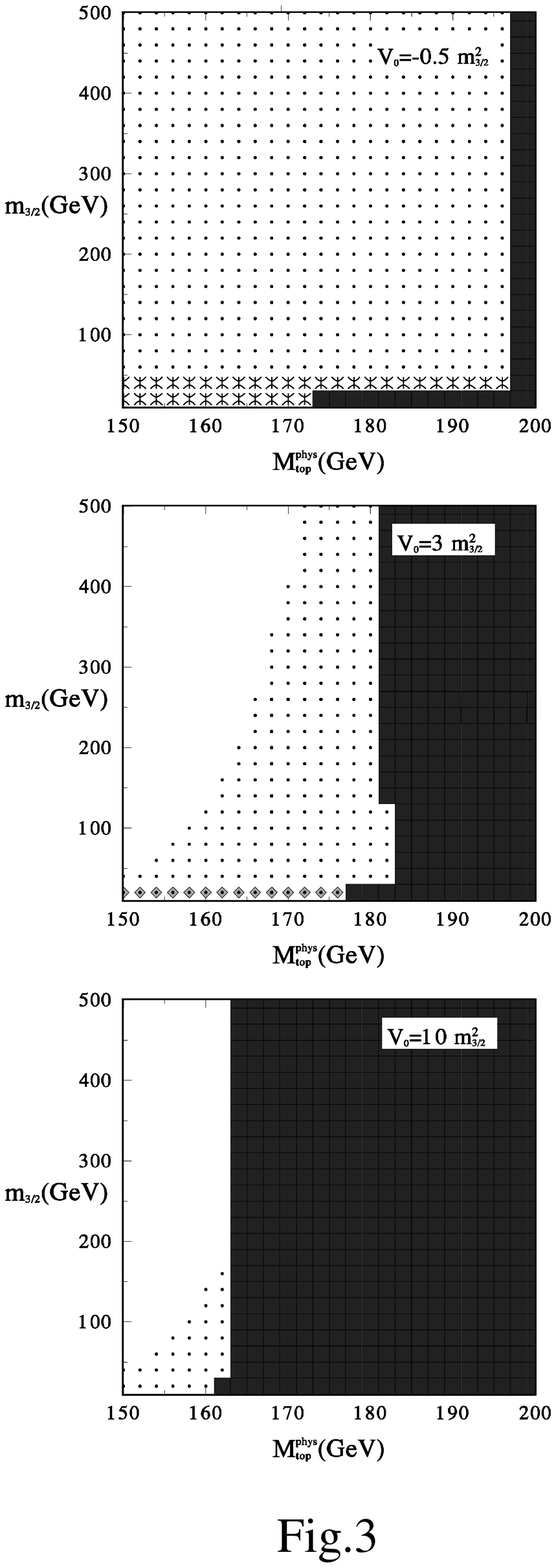,height=23.1cm,angle=180}
%
\end{document}